\documentclass[11pt]{article}
\usepackage{}
\usepackage{mathrsfs}

\renewcommand{\baselinestretch}{1.4}
  \renewcommand{\arraystretch}{1.2}
 \usepackage{amsmath}
\usepackage{amssymb}
\usepackage{graphicx}
\usepackage{ulem}
\usepackage{hyperref}

\begin{document}

 \title{A Note on ``New techniques for noninteractive zero-knowledge"}

  \author{Zhengjun Cao$^{1}$,  \  Lihua Liu$^{2,*}$}
  \footnotetext{ $^1$Department of Mathematics, Shanghai University, Shanghai, 200444,  China.  \\
   $^3$Department of Mathematics, Shanghai Maritime University, Shanghai, 201306,  China. \  $^*$\,\textsf{liulh@shmtu.edu.cn} }

 \date{}\maketitle

\begin{quotation}
\noindent \textbf{Abstract}.
  In 2012, Groth, et al. [J. ACM, 59 (3), 1-35, 2012] developed some new techniques for noninteractive zero-knowledge (NIZK) and presented:
    the first perfect NIZK argument system for all NP;  the first universally composable NIZK argument for all NP in the presence of an adaptive adversary;  the first noninteractive zap for all NP, which is based on a standard cryptographic security assumption. These solved several long-standing open questions.   In this note, we remark that their basic system is flawed because
   the prover can cheat the verifier to accept a false claim. Thus,
 these problems remain open now.

 \end{quotation}

\section{Introduction}

Non-interactive zero-knowledge (NIZK) proof in the common random
string model, introduced by Blum et al. \cite{Blum88},
 plays a key role in many constructions, including  digital signatures \cite{Garay03}, E-voting \cite{Groth05}, Shuffle \cite{Bayer12}, polynomial evaluation \cite{Bayer13}, arithmetic circuits \cite{Bootle162,Bootle161} and multiple-party computation  protocols.
 In 1988, Blum et al. \cite{Blum88}  constructed some computational NIZK proof systems
for proving a single statement about any NP language.
In 1991, they \cite{Blum91} presented the first computational
NIZK proof system for multiple theorems. These systems are  based on  the hardness of deciding quadratic
residues modulo a composite number.
 In 1998,   Kilian and Petrank \cite{Kilian98} designed an efficient noninteractive zero-knowledge proof
system for NP with general assumptions.

 In 1999, Feige et al. \cite{Feige99} developed a method  to construct computational NIZK proof systems based on any trapdoor permutation.
 Goldreich et al. \cite{Goldreich99} discussed the possibility of converting a statistical zero knowledge (SZK) proof into
a NIZK proof. In 2001,  Santis et al. \cite{Santis01,Santis02} investigated the robustness and randomness-optimal characterization of some NIZK proof systems.   In 2003,  Sahai and Vadhan \cite{Sahai03} presented an interesting survey on SZK.
Groth et al. \cite{Groth09,Groth10,GrothS12} designed some linear algebra with sub-linear zero-knowledge arguments and short pairing-based NIZK arguments.
In 2015, Gentry et al. \cite{Gentry15} investigated the problem of using fully homomorphic hybrid encryption to minimize NIZK proofs.

At EUROCRYPT'06, Groth, et al. \cite{GrothOS06} designed a new NIZK proof system for plaintext being 0 or 1 using bilinear groups with composite order. The refined version \cite{GrothOS12} was published by J. ACM  in 2012.
 The behind intractability is the subgroup decision problem introduced by Boneh et al. \cite{Boneh05}.
Based on the basic homomorphic proof commitment scheme, they presented: (1)
 the first perfect NIZK argument system for all NP;
 (2)  the first universally composable NIZK argument for all NP in the presence of an adaptive adversary;
(3)  the first noninteractive zap for all NP, which is based on a standard cryptographic security assumption.
 These solved several long-standing open questions.

In this note, we would like to remark that in their basic homomorphic proof commitment scheme the prover has not to invoke the trapdoor key to generate witnesses. The mechanism was dramatically different from the previous works, such as Blum-Feldman-Micali proof system \cite{Blum88} and Blum-Santis-Micali-Persiano proof system \cite{Blum91}. 
    We show  that the prover, who is accessible to the trapdoor key, can cheat the verifier to accept a false claim. Thus,
 these problems concerning NIZK protocols remain open now.

\section{Review of the basic homomorphic proof commitment scheme}

Let $\mathcal{G}_{\text{BGN}}$ be a randomized algorithm that on security parameter $k$ poutputs $(p, q, \mathbb{G}, \mathbb{G}_T, e, g)$ such that: (1) $p, q$ are primes with $p<q$; (2) $\mathbb{G}, \mathbb{G}_T$ are descriptions of cyclic groups of order $n=pq$; (3) $e: \mathbb{G}\times \mathbb{G}\rightarrow \mathbb{G}_T$  is a bilinear map; (4) $g$ is a random generator for $\mathbb{G}$ and $e(g, g)$ generates $\mathbb{G}_T$; (5) group operations, deciding group membership and the bilinear map are efficiently computable.
Let $\mathbb{G}_q$ be the subgroup of $\mathbb{G}$ of order $q$.

The homomorphic proof commitment scheme based on the subgroup decision assumption can be described as follow.

\noindent\rule[-0.25\baselineskip]{\textwidth}{0.25mm}

\noindent \textbf{Perfectly binding key generation}.  Generate $(p, q, \mathbb{G}, \mathbb{G}_T, e, g)\leftarrow \mathcal{G}_{\text{BGN}}(1^k)$ and set $n=pq$. Pick $x\leftarrow \mathbb{Z}_q^*$, and set $h=g^{px}$. Let
$ck=(n, \mathbb{G}, \mathbb{G}_T, e, g, h)$ and $xk=(ck, q)$.

\noindent \textbf{Perfectly hinding key generation}. Generate $(p, q, \mathbb{G}, \mathbb{G}_T, e, g)\leftarrow \mathcal{G}_{\text{BGN}}(1^k)$ and set $n=pq$. Pick $x\leftarrow \mathbb{Z}_q^*$, and set $h=g^{x}$. Let
$ck=(n, \mathbb{G}, \mathbb{G}_T, e, g, h)$ and $tk=(ck, x)$.

 \noindent \textbf{Commitment}. To commit to message $m\in \mathbb{Z}_p$, pick $r\leftarrow \mathbb{Z}_n$ and compute
 $c=g^mh^r$.

 \noindent \textbf{Extraction}. On a perfect binding key we can use $xk=(ck, q)$ to extract $m$ of length $\mathcal{O}(\log k)$ from $c=g^mh^r$ by computing $c^q=(g^mh^r)^q=(g^q)^m$ and exhaustively search for $m$.

 \noindent \textbf{Trapdoor opening}. Given a commitment $c=g^mh^r$ under a perfectly hiding commitment key we have
 $c=g^{m'} h^{r-(m'-m)/x}$. So we can create a perfectly hiding commitment and open it to any value we wish if we have the trapdoor key $tk=(ck, x)$. It returns $r'=r-\frac{m'-m}{x} \,\mathrm{mod}\,n$.

 \noindent \textbf{WI proof}.  Given $m, r\in \{0, 1\}\times \mathbb{Z}_n$ we make the WI proof for commitment to 0 or 1 as $\pi=(g^{2m-1}h^r)^r$.

 \noindent \textbf{Verification}. To verify a WI proof $\pi$ of commitment $c$ containing 0 or 1, check
 $$e(c, cg^{-1})=e(h, \pi)$$

  \noindent\rule[0.55\baselineskip]{\textwidth}{0.25mm}

It is easy to check its correctness because
  $$e(c, cg^{-1})=e(g^mh^r, g^{m-1}h^r)= \uwave{e(g, g)^{m(m-1)}}e(g, h)^{(2m-1)r} e(h, h)^{r^2}=\uwave{e(g, g)^{m(m-1)}} e(h, \pi)$$
  Since $m\in\{0, 1\}$, we have  $e(c, cg^{-1})=e(h, \pi)$.


\section{The proof commitment scheme is flawed}

\subsection{What is the true claim}

Give $c\in \mathbb{G}$, the prover claims that $c$ is of the form $g^mh^r$ for some $(m, r)\in \{0, 1\}\times \mathbb{Z}_n$.
This is equivalent to check whether $c$ or $c/g$ is in the subgroup $\mathbb{G}_q$.

If the trapdoor key $q$ is available to the verifier, then it suffices to check that $c^q=1$ or $(c/g)^q=1$.
However, the trapdoor key $q$ cannot be directly shown  to the verifier.  Therefore, the prover has to produce some witnesses to convince the verifier of that $c$ or $c/g$ is indeed in the subgroup $\mathbb{G}_q$.

\subsection{The prover can cheat the verifier to accept a false claim}

Notice that the system does not specify that who is responsible for generating the system parameters. If there is a third-party, Cindy,  who generates the system parameters, then Cindy is not fully trustable and she knows the trapdoor key. Otherwise,
\textit{the presence of a fully trustable third party is incompatible with the general model of zero-knowledge proof}.
 Therefore, the prover can form an alliance with Cindy.
All in all, the prover can access to the trapdoor key $(p, q)$. In this situation,  we now show that the prover can  cheat the verifier  to accept a false claim.

 The prover sets \underline{$c=g^{\alpha_1}h^{\alpha_2}, \pi=g^{\beta_1}h^{\beta_2}$}, where $\alpha_1, \alpha_2, \beta_1, \beta_2$ are to be determined.
 Since
$$e(c, cg^{-1})=e(g^{\alpha_1}h^{\alpha_2}, g^{\alpha_1-1}h^{\alpha_2})
=e(g, g)^{\alpha_1(\alpha_1-1)}e(g, h)^{\alpha_1\alpha_2+\alpha_2(\alpha_1-1)}e(h, h)^{\alpha_2^2}$$
$$e(h, \pi)=e(h, g^{\beta_1}h^{\beta_2})=e(g, h)^{\beta_1}e(h, h)^{\beta_2} $$
it suffices for the prover to solve
$$ \left\{
\begin{array}{l}
\alpha_1(\alpha_1-1)= 0 \,\mathrm{mod}\,n  \\
2\alpha_1\alpha_2-\alpha_2=\beta_1 \,\mathrm{mod}\,n \\
\alpha_2^2=\beta_2   \,\mathrm{mod}\,n
\end{array} \right. $$
for those exponents.

Armed with the trapdoor key $p, q$, the prover computes $k, \ell$ using Extended Euclid Algorithm such that
$$kq-\ell p=1.$$
He then sets \underline{$\alpha_1=kq$}, picks
\underline{$\beta_1<n$} and computes \underline{$\alpha_2=\beta_1(2kq-1)^{-1}\,\mathrm{mod}\,n, \beta_2=\alpha_2^2 \,\mathrm{mod}\,n $}.

It is easy to check that the above values $c, \pi$ pass the verification. 
Clearly, $\alpha_1=kq\neq 0, 1$.
Besides, 
$$(g^{\alpha_1})^q=(g^{kq})^q=(g^{\ell p+1})^q=g^{q}\neq 1$$ 
namely $g^{\alpha_1}\not\in \mathbb{G}_q$. Therefore, since $h^q=(g^{px})^q=1$, there does not exist an integer $\alpha'$ such that $g^{\alpha_1}=h^{\alpha'}$.
That means $c=g^{\alpha_1}h^{\alpha_2}$ cannot be eventually expressed as $h^{w_1}$ or $gh^{w_2}$ for some integers $w_1, w_2$. Thus, the prover can cheat the verifier to accept a false claim.

\section{Conclusion}

We present an attack against the basic homomorphic proof commitment scheme proposed by Groth et al. in 2012. 
The system seems secure if the trapdoor key is indeed not accessible to the prover. But the assumption that the presence of a fully trustable third party is somewhat incompatible with the general primitive of zero-knowledge proof, and makes the system itself unsuitable to more broader applications.


\end{document}